\documentclass[aps,prx,twocolumn,superscriptaddress]{revtex4-2}

\usepackage{amsmath,amssymb}
\usepackage{graphicx}
\usepackage{epstopdf}
\usepackage{bm,ulem}
\usepackage{xcolor}

\newcommand{\co}	{$^{59}$Co}
\newcommand{\slr} 	{$T_1^{-1}$}
\newcommand{\ssr} 	{$T_2^{-1}$}
\newcommand{\ssL} 	{$T_{2L}^{-1}$}
\newcommand{\ssG} 	{$T_{2G}^{-1}$}
\newcommand{\slrt} 	{$(T_1T)^{-1}$}

\begin{document}

\title{Possible quadrupole-order-driven commensurate-incommensurate phase transition in B20 CoGe}

\author{S.-H. Baek}
\email[]{sbaek.fu@gmail.com}
\affiliation{Department of Physics, Changwon National University, Changwon 51139, Korea}
\affiliation{Department of Materials Convergence and System Engineering, Changwon National University, Changwon 51139, Korea}
\author{V.~A. Sidorov}
\affiliation{Vereshchagin Institute for High Pressure Physics, RAS, 108840 Moscow, Troitsk, Russia}
\author{A.~V. Nikolaev}
\affiliation{Skobeltsyn Institute of Nuclear Physics Lomonosov Moscow State University, 119991 Moscow, Russia}
\author{T. Klimczuk}
\affiliation{Faculty of Applied Physics and Mathematics, Gdansk University of Technology, Narutowicza 11/12, 80-233 Gdansk, Poland}
\affiliation{Advanced Materials Centre, Gdansk University of Technology, Narutowicza 11/12, 80-233 Gdansk, Poland}
\author{F. Ronning}
\affiliation{Los Alamos National Laboratory, MPA-CMMS, Los Alamos, New Mexico 87545, USA}
\author{A.~V. Tsvyashchenko}
\affiliation{Vereshchagin Institute for High Pressure Physics, RAS, 108840 Moscow, Troitsk, Russia}

\date{\today}


\begin{abstract}

The B20-type cobalt germanide CoGe was investigated by measuring the specific heat, resistivity, and \co\ nuclear magnetic resonance (NMR). We observed a phase transition at $T_Q=13.7$ K, evidenced by a very narrow peak of the specific heat and sharp changes of the nuclear spin-spin (\ssr) and spin-lattice (\slr) relaxation rates.  The fact that the entropy release is extremely small and the Knight shift is almost independent of temperature down to low temperatures as anticipated in a paramagnetic metal indicates that the $T_Q$ transition is of non-magnetic origin.
In addition, we detected a crossover scale $T_0\sim30$ K below which the resistivity and the NMR linewidth increase, and \slr\ is progressively distributed in space, that is, a static and dynamical spatial inhomogeneity develops. While the order parameter for the $T_Q$ transition remains an open question, a group-theoretical analysis suggests that the finite electric quadrupole density arising from the low local site symmetry at cobalt sites could drive 
the crystal symmetry lowering from the P2$_1$3 symmetry that is commensurate to the R3 symmetry with an incommensurate wavevector, which fairly well accounts for the $T_Q$ transition. 
The quadrupole-order-driven commensurate-incommensurate phase transition may be another remarkable phenomenon arising from the structural chirality inherent in the noncentrosymmetric B20 family.

\end{abstract}

\maketitle


\section{Introduction}

Transition-metal mono-silicides $M$Si and -germanides $M$Ge with the B20 cubic structure, where $M$ denotes a transition metal such as Mn, Fe, and Co, are noncentrosymmetric materials carrying inherent structural chirality (lack of mirror symmetry) which has a drastic influence on the electronic and magnetic properties of these materials, being of recent special interest from both fundamental and technical points of view \cite{pshenayseverin19}.
On one hand, the structural chirality causes finite spin-orbit-driven Dzyaloshinskii-Moriya interaction, which competes with the Heisenberg exchange coupling. As a result, a helical spin ordered state whose chirality is determined by the structural one \cite{grigoriev10}, is stabilized as observed in MnSi \cite{ishikawa76,stishov11}, FeGe \cite{lebech89}, MnGe \cite{kanazawa11,makarova12,chtchelkatchev19}, and Fe$_{1-x}$Co$_x$Si \cite{beille81}. It has been shown that the application of a small magnetic field in these materials generates peculiar particle-like localized spin textures, such as magnetic skyrmion lattices in MnSi \cite{muehlbauer09,nakajima17} and Fe$_{1-x}$Co$_x$Si \cite{muenzer10,yu10}, a topological spin texture in MnGe \cite{fujishiro18}, and chiral magnetic bobbers in FeGe \cite{zheng18}. Furthermore, under hydrostatic pressure, chiral magnets undergo a quantum phase transition into a non-Fermi liquid for MnSi \cite{pfleiderer04,ritz13} or an inhomogeneous chiral-spin state for FeGe \cite{barla15}.
On the other hand, for nonmagnetic systems, e.g., CoSi and RhSi, the structural chirality results in a three-dimensional chiral topological semimetal featured by unconventional chiral fermions \cite{ishii14,tang17,takane19,rao19,sanchez19,xu20a} and a circular photo-galvanic effect \cite{chang17,ni21}. Interestingly, chiral magnetism with skyrmionic spin texture could be achieved in Co$_{1+x}$Si$_{1-x}$ alloys for $x\geq0.028$ \cite{balasubramanian20}.

\begin{figure*}
\centering
\includegraphics[width=\linewidth]{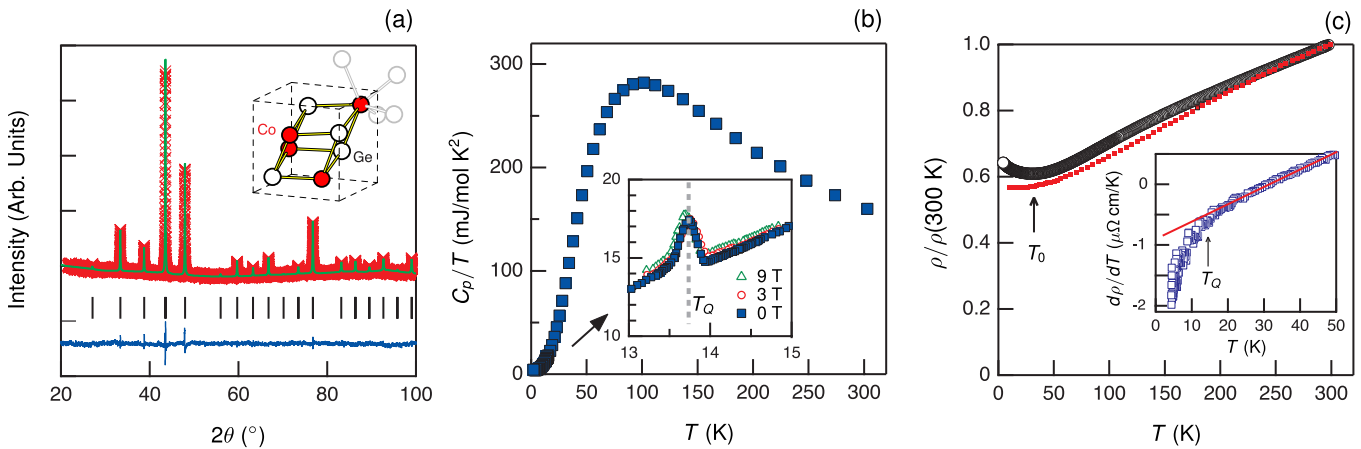}
\caption{(a) X-ray powder diffraction pattern of CoGe measured at $T = 295$ K at ambient pressure. The experimental points (red marks), the calculated profile (green line) and their difference (blue line) are shown. The black bars in the lower part of the graph represent the calculated Bragg reflections that correspond to the B20 structure. (b) The specific heat divided by temperature of CoGe between 2 K and 300 K. The inset enlarges the data near the sharp peak found at $T_Q=13.7$ K. The peak is almost robust against the magnetic field up to 9 T. (c) Temperature dependence of the resistivity of CoGe, normalized at 300 K, which forms a minimum at $T_0\sim30$ K. The dotted data were taken from DiTusa \textit{et al.}\,\cite{ditusa14}.  The inset shows $d\rho(T)/dT$ versus $T$ at low temperatures.
}
\label{structure}
\end{figure*}

While the structural chirality is an essential feature of the B20 binary compounds, the electron count in the $3d$ electron shell of the transition element $M$ is another crucial factor for the ground state. On the other hand, the structural chirality should have a considerable effect on an unfilled $3d$ shell of $M$, regardless of the detailed valence state which is not well-defined in the B20 alloys (for example, see \cite{carbone06}). That is, since the local site symmetry at $M$ is quite low in the B20 structure, the partial occupation of the $3d$ shell implies that the density matrix elements $\langle\psi_{lm}|\rho_{L,M}|\psi_{l'm}\rangle$ ($l = l' = 2$ for $d$-functions) have a significant quadrupole electron density at corresponding crystal sites. The quadrupole contributions could arise also from the mixture of $3d$ and $4s$ functions, $\langle\psi_{lm}|\rho_{L,M}|\psi_{l'=0}\rangle$ where the function $\psi_{l'=0}$ refers to the $4s$ states \cite{bradley72}. 
The appreciable quadrupole density can lead to strong anisotropic electronic interactions which can drive a structural or quadrupole phase transition \cite{bradley72,santini09,nikolaev01}.
Such a situation is not uncommon for $4f$- and $5f$-elements, where a quadrupole moment is caused by mixing of $f$-states \cite{santini09,nikolaev01}, as observed in, for example, CeB$_6$ ($T_Q=3.3$ K) \cite{santini09,mannix05}, TmTe ($T_Q=1.8$ K) \cite{matsumura98}, and NpO$_2$ ($T_Q = 25.5$ K) \cite{santini09}.
Quadrupole phase transitions are, however, quite rare in $3d$ based compounds.

In this paper, we carried out an experimental and theoretical investigation in B20 CoGe, one of the least studied B20 compounds partly due to the difficulty in synthesizing B20-type CoGe which requires the application of high pressure and temperature \cite{larchev82, takizawa88}. So far, B20 CoGe has been known as a Pauli paramagnet with a low carrier density without a phase transition \cite{tsvyashchenko12,kanazawa12,ditusa14}, similar to the silicide counterpart CoSi, while being distinguished from other close germanides FeGe and MnGe that are helimagnets. In fact, the band crossings in the $\Gamma$ and $R$ points of the Brillouin zone in CoGe \cite{kanazawa12,ditusa14} bear remarkable resemblance to those in CoSi \cite{rao19}, suggesting that CoGe could host unconventional chiral fermions.

While as yet little was known of a phase transition other than the magnetic one in the B20 family, our experimental findings, such as anomalous changes of the \co\ nuclear relaxation rates and a very sharp peak in the specific heat, evidence the occurrence of a non-magnetic phase transition at $T_Q=13.7$ K in CoGe. By performing a group-theoretical analysis, we propose that the $T_Q$ phase transition may be a commensurate-incommensurate structural transition driven by electric quadrupole density which arises from the structural chirality.
Our experimental data further suggest that some spatial inhomogeneity develops at a higher temperature $T_0\sim30$ K than $T_Q$.


\section{Experimental details}

Polycrystalline samples of B20-cubic CoGe were synthesized under a pressure of 8 GPa and at high temperatures of 1500--1700 K using a toroidal high pressure apparatus \cite{khvostantsev04} by melting Co and Ge \cite{tsvyashchenko84}. The purity was 99.9\% for Co, and 99.999\% for Ge. The B20 phase of high pressure synthesized samples remains metastable after the pressure release at room temperature, while the monoclinic phase appears only after heating the sample above 1000 K at ambient pressure. The crystal structure  was examined by x-ray diffraction (XRD), as shown in Fig.~1(a). Measurements were performed at room temperature and ambient pressure using the diffractometer STOE IPDS-II (Mo $K\alpha$) and Guinier camera e G670, Huber (Cu $K\alpha1$). Powder x-ray diffraction revealed a single phase material with cubic B20-type structure. The lattice parameter of CoGe (0.46392 nm) is close to those published earlier (0.4637 nm \cite{larchev82} and 0.4631 nm \cite{takizawa88} for CoGe). In the B20 structure both Co and Ge are located at the Wyckoff positions (4a) with coordinates $(u, u, u)$, $(u + 0.5, 0.5 - u, -u)$, $(-u, 0.5 + u, 0.5 -u)$, and $(0.5- u, -u, 0.5 +u)$. 
Our refinement determined $u$ to be 0.133 for Co and 0.838 for Ge. Also, the Rietveld fit of the x-ray powder diffractogram shows that the site population ratio Co/Ge is 0.999(5), evidencing a negligibly small intermixing between the two ions.

Measurements of the specific heat $C_p$ were performed using the Quantum Design Physical Properties Measurement System (PPMS). Resistivity $\rho$ was measured by a four-terminal method using spot welded Pt wires as the electrodes.

\co\ (nuclear spin $I=7/2$) NMR experiments were carried out at the external magnetic field of $H=3$ T in the temperature range $T=2.5-200$ K. The NMR spectra and the spin-spin relaxation rate \ssr\ were obtained using the Hahn echo sequence with a typical $\pi/2$ pulse length of $\sim3 \;\mu$s. For the full spectra including satellite transitions, we used a frequency-sweep method.  The spin-lattice relaxation rate \slr\ was measured by the saturation method on the first satellite line and determined by fitting the recovery curve of the nuclear magnetization $M(t)$ to the appropriate fitting function for the first satellite transition in the $I=7/2$ case with the inclusion of a stretching exponent $\beta$, i.e., 
\begin{equation}
\begin{split}
\label{eq:T1} \textstyle
1-\frac{M(t)}{M(\infty)}=&
 \textstyle A\left[\frac{1}{84}e^{-(t/T_1)^\beta}+\frac{1}{84}e^{-(3t/T_1)^\beta} + \frac{1}{33}e^{-(6t/T_1)^\beta}\right.   \\
&+  \textstyle \frac{9}{77}e^{-(10t/T_1)^\beta}+\frac{1}{1092}e^{-(15t/T_1)^\beta} \\
&+ \textstyle \left.\frac{49}{132}e^{-(21t/T_1)^\beta} + \frac{196}{429}e^{-(28t/T_1)^\beta}\right],
\end{split}
\end{equation} 
where $A$ is a fitting parameter that is ideally one.

\begin{figure*}
\centering
\includegraphics[width=\linewidth]{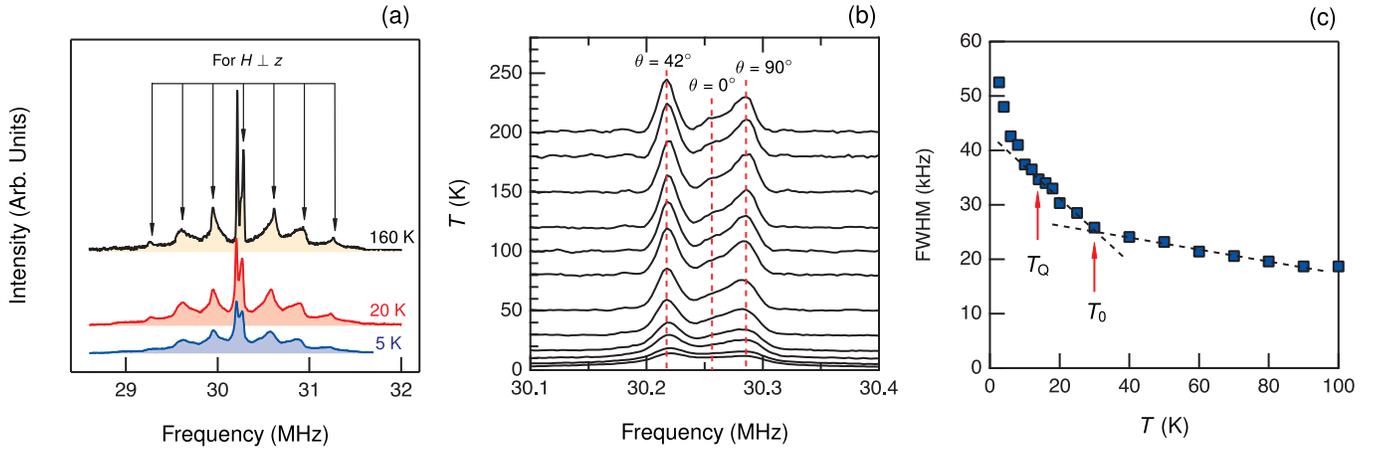}
\caption{(a) \co\ NMR spectra obtained at 3 T by a frequency sweep. They show a typical spectral pattern found in a randomly oriented powder for $I=7/2$ nuclei with near axial symmetry. The vertical arrows are expected lines for $H\perp z$ ($\theta=90^\circ$) in the absence of asymmetry ($\eta=0$). With lowering temperature, the spectrum is moderately broadened with a small reduction of the separation of satellites. (b) The spectrum at the central transition which is determined by the second order quadrupole shift as a function of temperature. The two maxima as well as the $\theta=0^\circ$ position are nearly unshifted with lowering temperature, indicating that the Knight shift is unchanged. (c) Temperature dependence of the FWHM of the lower central peak in (b). Anomalous increase of the FWHM at $T_0\sim30$ K is noticed. The dotted lines are a guide to the eyes to emphasize the anomaly at $T_0$.
}
\label{spec}
\end{figure*}

\section{Results}

\subsection{Specific heat and resistivity}

The specific heat divided by temperature, $C_p/T$, as a function of temperature is shown in Fig.~1(b). While the data are consistent with the previous results reported in the literature \cite{tsvyashchenko12, ditusa14}, we detected a weak but very sharp peak at $T_Q= 13.7$ K, as shown in the inset of Fig.~1(b), which has been overlooked so far. 
There are two important features of this specific heat peak that are prominent: (i) the peak position and amplitude are nearly independent of external magnetic field strength up to 9 T, and (ii) an extremely small magnetic entropy is released around the transition [$\Delta S = 0.89$ mJ/mol K = 0.00015($R\ln2$)]. Because the uniform magnetic susceptibility $\chi(T)$ does not exhibit any anomaly near $T_Q$ \cite{tsvyashchenko12}, these features indicate that the transition is of non-magnetic origin.  Below, we will show that the low energy spin/charge dynamics probed by the \co\ relaxation rates exhibit clear anomalous changes at $T_Q$, while the Knight shift, or the local static spin susceptibility, is unchanged through $T_Q$, demonstrating that the $T_Q$ transition is an intrinsic bulk property of the material.

Resistivity measurements were performed on a few samples synthesized at high pressure. General behavior of $\rho(T)$ for all samples is similar, while they differ in the value of the residual resistivity. Representative $\rho(T)/\rho(\text{300 K})$ data of CoGe are shown in Fig.~1(c). We also compared the data from DiTusa \textit{et al.}~\cite{ditusa14}.
Metallic behavior in $\rho(T)$ is observed at high temperatures between 30 and 300 K. However, our $\rho(T)$ data reveal a distinguished feature at low temperatures : $\rho(T)$ passes through a minimum at a characteristic temperature $T_0\sim30$ K and is slightly enhanced upon further cooling. It appears that $\rho(T)$ is quite symmetric around $T_0$, as evidenced by the linear temperature derivative of the resistivity, $d\rho(T)/dT$, between $T_Q$ and 50 K [see red line in the inset of Fig.~1(c)]. The enhancement of $\rho(T)$ at low temperatures resembles the Kondo effect, but it should be noted that the $\gamma$-value estimated from the $C/T$ data is only about 3 mJ/mol K$^2$, which is comparable to the free electron value. Also, there is no signature of magnetic screening from the susceptibility measurements \cite{tsvyashchenko12,ditusa14}. Therefore, the enhanced resistivity below $T_0$ may be related with a spatial inhomogeneity as discussed below. 

\subsection{\co\ NMR spectrum and Knight shift}

Figure \ref{spec}(a) shows the quadrupole-perturbed \co\ spectra in CoGe obtained by a frequency sweep method at 3 T at three selected temperatures.
For a nuclear spin $I> 1/2$ in an axial symmetric surrounding (asymmetry parameter $\eta=0$), to first order, there are central ($-\frac{1}{2}\leftrightarrow \frac{1}{2}$) and satellite transitions between the $m$th and $(m-1)$th levels ($m=-I,-I+1,\cdots, +I$) which are given by
\begin{equation}
	\begin{split}
\nu(m\leftrightarrow m-1)  =\; &\nu_0 (1+\mathcal{K}) \\
&+\frac{1}{2}\nu_Q(3\cos^2\theta -1)\left(m-\frac{1}{2}\right),
	\end{split}
	\label{quad1}
\end{equation}
where $\mathcal{K}$ is the Knight shift, $\nu_0$ is the unshifted Larmor frequency, $\nu_Q$ is the nuclear quadrupole frequency, and $\theta$ is the angle between the principal axis $z$ of the electric field gradient (EFG), which is most likely along the diagonals,  and an external field $H$.

The observed full NMR spectrum is roughly described by Eq.\,(\ref{quad1})  in a randomly oriented powder, in which the $m\leftrightarrow m-1$ transitions for $\theta=90^\circ$ ($H\perp z$) form the sharp satellite peaks which are equally spaced by $\nu_Q/2$, indicating that the local symmetry at \co\ is close to axial.
Nevertheless, we find that the positions of two outer pairs of satellites slightly differ from the expected ones for $\eta=0$, implying a finite asymmetry. In this case, the distance between the first satellites is equivalent to the \textit{apparent} quadrupole frequency $\nu_Q'$ 
that is related to the actual value by $\nu_Q=\nu_Q'/(1-\eta)$ \cite{bennet}.

In addition to the first order quadrupole effect, there is also a second order quadrupole effect which shifts the central transition ($-\frac{1}{2}\leftrightarrow \frac{1}{2}$) depending on $\theta$. For $I=7/2$, it is written :
\begin{equation}
\label{quad2}
	\begin{split}
	\nu ({\textstyle-\frac{1}{2}\leftrightarrow \frac{1}{2}}) =& \nu_0 (1+\mathcal{K}) \\ &+ \frac{15\nu_Q^2}{16\nu_0}(1-\cos^2\theta)(1-9\cos^2\theta).
	\end{split}
\end{equation}
The powder pattern of the central transition as a function of temperature is presented in Fig.~\ref{spec}(b). It reveals two peaks which correspond to the lines for $\theta\sim 42^\circ$ and $90^\circ$ at which the second order quadrupole shift is minimal and maximal, respectively, as can be easily verified in Eq.\,(\ref{quad2}). The separation between the two maxima $\Delta\nu$ depends on $\eta$ as well as $\nu_Q$ \cite{bennet}:
\begin{equation}
\label{deltanu}
	\Delta\nu = \frac{15\nu_Q^2}{144\nu_0}(\eta^2+22\eta+25),
\end{equation}
where $\nu_Q=\nu_Q'/(1-\eta)$. Since $\Delta\nu$ is only slightly decreased at low temperatures without a noticeable anomaly, we compare the $\nu_Q$ and $\eta$ values at the two temperatures, 160 K and 5 K, at which the full spectra are available. By using the $\nu_Q'$ values extracted from the full spectra in Fig.~2(a), Eq.~(\ref{deltanu}) yields $\eta=0.191$ at 160 K and $\eta=0.222$ at 5 K which, in turn, give rise to $\nu_Q=0.821$ MHz at 160 K and 0.793 MHz at 5 K. 
This shows that the asymmetry is indeed finite reflecting the low local site symmetry at \co\ with weak temperature dependence.

Unlike the two sharp maxima for the central transition, the $\theta=0^\circ$ line, which often appears as a step-like anomaly in a clean sample as observed at high temperatures [see the dotted line for $\theta=0^\circ$ in Fig.~\ref{spec}(b)], is unaffected by quadrupole effects up to second order, and thus permits us to deduce the Knight shift $\mathcal{K}$. Although the step-like anomaly gradually smears out at low temperatures due to line broadening, the data indicate that the Knight shift is essentially independent of temperature down to 2.5 K, evidencing that the system remains a Pauli paramagnet regardless of the $T_Q$ phase transition.

Another interesting feature found in Figs.~\ref{spec}(a) and \ref{spec}(b) is the broadening of the NMR spectrum at low temperatures. As shown in Fig.~\ref{spec}(c), the full width at half maximum (FWHM) measured on the left central peak for $\theta=42^\circ$ is more rapidly enhanced below $T_0\sim30$ K, suggesting that a static spatial inhomogeneity develops below $T_0$ and becomes even stronger at low temperatures toward zero temperature.

\begin{figure}
\centering
\includegraphics[width=0.75\linewidth]{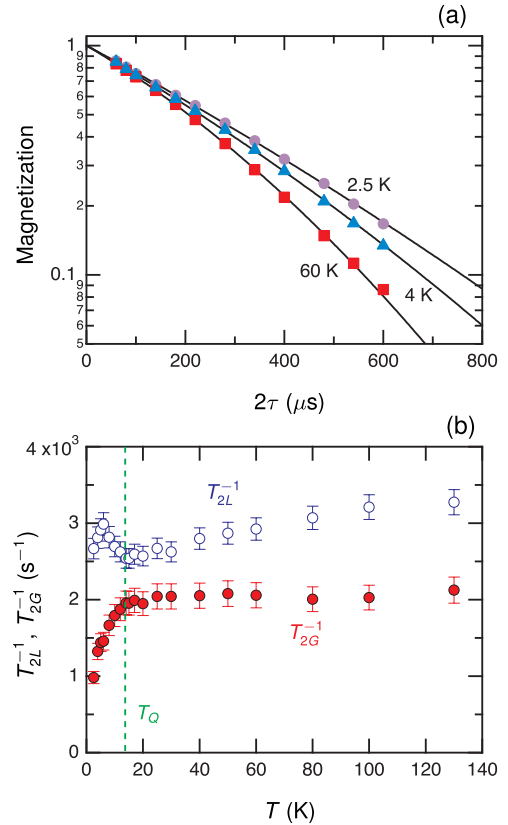}
\caption{(a) Decay of the nuclear magnetization $M_\text{n}$ of \co\ with time $2\tau$ at selected temperatures. At low temperatures, the Lorentzian rate \ssL\ becomes dominant over the Gaussian one \ssG. (b) Temperature dependence of \ssL\ and \ssG. The sharp anomalous changes of both rates are observed at $T_Q=13.7$ K.  }
\label{t2}
\end{figure}

\subsection{\co\ nuclear relaxation rates}

\begin{figure}
\centering
\includegraphics[width=0.75\linewidth]{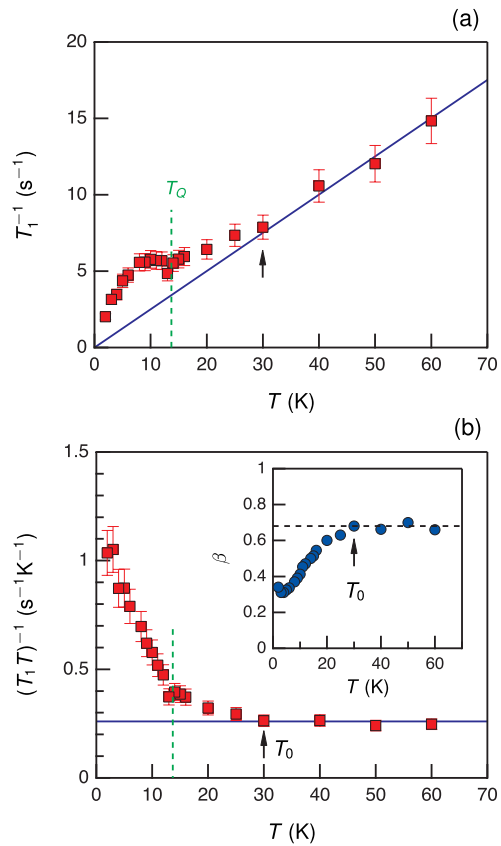}
\caption{Temperature dependence of (a) \slr\ and (b) \slrt.  The data deviate from the linear $T$-dependence roughly below $T_0\sim30$ K and reveal a small kink at $T_Q=13.7$ K followed by a rapid increase of \slrt. The inset shows the temperature dependence of the stretching exponent $\beta$, which is a measure of a dynamical spatial inhomogeneity. }
\label{t1}
\end{figure}

Figure \ref{t2}(a) shows the decay of the nuclear magnetization $M_\text{n}$ of \co\ at three selected temperatures, being normalized at zero time. $M_\text{n}$ is described by both the Lorentzian and Gaussian spin-spin relaxation rate, \ssL\ and \ssG :
\begin{equation}
	M_\text{n}(2\tau) = M_0 \exp\left(-\frac{2\tau}{T_{2L}}\right)\exp\left[-\frac{1}{2}\left(\frac{2\tau}{T_{2G}}\right)^2\right],
\end{equation}
where $\tau$ is the delay time between the echo pulses. In general, the two rates result from different relaxation processes : while \ssL\ is the contribution from the spin-lattice relaxation process, \ssG\ describes the indirect coupling between neighboring nuclear spins through electronic excitations \cite{pennington91}. 
Clearly, as shown in Fig.\,2(b), both \ssL\ and \ssG\ significantly change at $T_Q=13.7$ K. Below $T_Q$, \ssL\ is enhanced forming a peak centered at $\sim 7$ K. One can see that its temperature dependence is quite similar to that of \slr\ [see Fig.~\ref{t1}(a)], supporting the direct relationship between \ssL\ and \slr.  In contrast, the Gaussian contribution \ssG\ is nearly temperature independent at high temperatures, but  rapidly decreases with decreasing temperature below $T_Q$, contrasting sharply with the upturn of \ssL.

The measurement of the spin-lattice relaxation rate \slr\ was carried out in the relatively low temperature region $T\leq 60$ K. The results are shown in Fig.\,\ref{t1}(a).
\slr\ is proportional to $T$ down to 30 K, or the constant $(T_1T)^{-1}$, as shown in Fig.~\ref{t1}(b). Together with the $T$-independent Knight shift, this indicates that the so-called Korringa law $(T_1T)\mathcal{K}^2 = \text{const.}$, which is expected in a normal metal, holds. Below $T_0\sim30$ K, however, the Korringa behavior breaks down. At the same time, we find that the stretching exponent $\beta$ [see Eq.\,(\ref{eq:T1})], shown in the inset of Fig.\,\ref{t1}(b), decreases from a constant below $T_0$, indicating that \slr\ is progressively distributed in space. 
Remarkably, the temperature at which $\beta$ begins to decrease is very close to $T_0$ at which the resistivity $\rho(T)$ forms a minimum [see Fig.~1(c)] and the NMR linewidth increases more rapidly [see Fig.~\ref{spec}(c)]. These findings strongly suggest that $T_0\sim 30$ K indeed represents a crossover scale below which a both \textit{static and dynamical} spatial inhomogeneity of electronic phases begins to develop. Also, we note that $\beta$ is a constant but still notably smaller than unity above $T_0$, suggesting that a temperature-independent weak spatial inhomogeneity already exists at high temperatures.

Upon further cooling, we find an additional increase of \slrt\ just below $T_Q$. Since the system remains a paramagnetic metal down to very low temperatures, the increase of \slrt\ below $T_Q$ suggests that the $T_Q$ transition is intimately related to a slowing of collective spin/charge fluctuations. Regardless of the underlying mechanism, the clear changes of the nuclear relaxation rates, \ssL, \ssG, and \slrt\ at $T_Q$ demonstrate that the $T_Q$ transition is the intrinsic bulk property of B20 CoGe.

\section{Discussion}


%

Combining our experimental findings, it is clear that a spatially inhomogeneous phase appears below a crossover scale $T_0\sim 30$ K and a non-magnetic long-range phase transition occurs at $T_Q=13.7$ K in B20 CoGe.
In order to obtain further insights into the possible order parameter for the $T_Q$ transition and the underlying origin of the spatial inhomogeneity developed below $T_0$, we employ group theory methods as follows.

As the B20 structure with the P2$_1$3 space group symmetry ($\Gamma_c$ T$^4$, \#198) lacks inversion symmetry, the local site symmetry in CoGe is quite complicated. There are four different 3-fold rotational axes directed along the cube diagonals, but they intersect at different cobalt and germanium atoms.
This symmetry imposes a certain restriction on the electron density component expressed in terms of spherical harmonics $Y_L^M$ about atoms. In particular, the expansion of electron density at cobalt sites has the $Y_{L=2}^0$ quadrupole ($L = 2$) and even the $Y_{L=1}^0$ dipole ($L = 1$) component lying along the cube diagonals \cite{bradley72}.

%
An electric quadrupole interaction  can lead to an effective attractive interaction between electrons at a high symmetry point of the Brillouin zone \cite{nikolaev12}. The effective interaction between electrons  can lower the crystal energy by driving a structural phase transition from P2$_1$3 to a phase with lower space symmetry. This implies a breaking of equivalence between four 3-fold rotational axes.
In this case, there is only one relatively high symmetry for the quadrupole ordering: the R3 space symmetry ($\Gamma_\text{rh}$ $C_3^4$, \#146). In R3 symmetry, there is only one 3-fold rotational axis (e.g., [111] in terms of cubic system) and there are two different channels for the P2$_1$3 $\rightarrow$ R3 symmetry lowering \cite{stokes88}. The first involves the quadrupole active mode at the $X$ point [$q_X=\pi/a (1,0,0)$] of the Brillouin zone, while the second is connected with the quadrupole mode at the $R$ point [$q_R=\pi/a (1,1,1)$]. Which mode actually drives the transformation depends on the energy lowering associated with the mode and requires an additional study. [Both modes ($X$ and $R$) and the related expansion of the Landau free energy are described in detail in the Appendix.]

As described in the Appendix, the theory shows that the \textit{Lifshitz condition} for both $q_X$ and $q_R$ modes fails, i.e., in the expansion of the Landau free energy in powers of the order parameter components $\eta_i$, the antisymmetrical terms, $\eta_k \frac{\partial\eta_i}{\partial x}-\eta_i\frac{\partial\eta_k}{\partial x}$ etc, are not zero \cite{lifshitz80}. Consequently, compared to the high-temperature P2$_1$3 symmetry that is commensurate, the low-temperature phase is expected to be of the R3 symmetry but with an incommensurate wave vector $q$ lying close to $q_X$ or $q_R$.
That is, both discussed schemes of the P2$_1$3 $\rightarrow$ R3 symmetry lowering is a commensurate-incommensurate phase transition induced by quadrupole ordering. 
Note that one should expect a weak anomaly of the specific heat at the transition because  the transformation due to the gradient contribution to the free energy is smoother, which is indeed the case at $T_Q$ in the specific heat measurement [see the inset of Fig.~1(b)].
In this case, the weak anomalous changes of low energy spin/charge dynamics at $T_Q$, probed by all the \co\ relaxation rates (\ssL, \ssG, and \slr), may be understood by the coupling between the quadrupole electron density and the spin moment.

The static and dynamical spatial inhomogeneity below $T_0\sim 30$ K, as inferred from the reduction of the stretching exponent $\beta$, the upturn of the resistivity $\rho(T)$, and the additional NMR line broadening, may be understood within the quadrupole ordering scenario, if the incommensurate modulation of the quadrupole electron density exists even above $T_Q$ which could cause a spatial inhomogeneity.
It may be worthwhile to recall that an intermediate chiral fluctuating region exists above $T_N$ in MnGe \cite{deutsch14}, which could be explained by the peculiar band structure.


\section{Summary}

We have established that the  sharp specific heat peak with a negligibly small release of the entropy at $T_Q=13.7$ K
represents a non-magnetic phase transition, an unprecedented one in the isostructural B20 family of materials. The bulk collective nature of the phase transition  at $T_Q$ was verified by probing the clear changes of the \co\ spin-spin and spin-lattice relaxation rates.

The application of group theory suggests that quadrupole electron density arising from the low local site symmetry of the transition metal element in the B20 structure could drive a structural symmetry lowering from P2$_1$3 to R3. For this symmetry transformation, the Lifshitz condition is violated and therefore the low-temperature R3 phase should be incommensurate. Indeed, the commensurate-incommensurate structural phase transition accounts for the tiny sharp peak in the specific heat observed at $T_Q$.
Whereas the (long-range) phase transition takes place at $T_Q$, the resistivity, the stretching exponent, and the NMR linewidth data point that a static and dynamical spatial inhomogeneity develops at a crossover scale $T_0$ considerably above $T_Q$.

It should be emphasized that our group-theoretical arguments could be almost equally applied in the other B20 compounds, and thus the quadrupole-order-driven commensurate-incommensurate phase transition may be ubiquitous in them. However, the tiny anomaly in the specific heat could be easily missed or overwhelmed by other effects, explaining why such a phase transition has hardly been found in these compounds despite many decades of study.


\begin{acknowledgments}
We thank D.A. Salamatin for the analysis of x-ray powder diffractograms.
This work was supported by the National Research Foundation of Korea (NRF) grant funded by the Korea government(MSIT) (Grant No. NRF-2020R1A2C1003817). T.K. was supported by the National Science Centre (Poland; Grant UMO-2018/30/M/ST5/00773).
Work at Los Alamos National Laboratory was performed under the auspices of the US DOE, Office of Basic Energy Sciences, Division of Material Sciences and Engineering.
\end{acknowledgments}


\bibliography{mybib}

\appendix*\section{Group-theoretical analysis}

The $X$ point of the B20 Brillouin zone has three rays [at $q_{X,1}=2\pi/a(1,0,0), q_{X,2}=2\pi/a(0,1,0), q_{X,3}=2\pi/a(0,0,1)$], and the active mode belongs to the two-dimensional irreducible representation $E$ of the little group of $X$. Therefore, the P2$_1$3 space group irreducible representation at the $X$ point (the $X_1$ irreducible representation) has six components (two components for each ray). A possible mechanism for symmetry lowering from P2$_1$3 to R3 involves  condensation of the component in the form ($a,b,a,b,a,b$) in the six-dimensional space of $X_1$ \cite{stokes88}. Thus, there are two different order parameter amplitudes: $a$ and $b$. Condensation of a single component of one ray [for example at $2\pi/a(1,0,0)$] implies that this component changes sign in real space in going from one crystallographic plane (perpendicular to the $x$ axis) to another. Condensation of all three components means that the sign changes in going from one plane to another along the $x$, $y$, and $z$ axes. Such structures are called triple-$q$-antiferroquadrupolar (3-q-AFQ) \cite{nikolaev12}. In real space the new quadrupole electron density is given by
\begin{equation*}
	\begin{split}
	\rho(\mathbf{R}_n,\mathbf{r}) &=  a \left( \rho_{1x}(\mathbf{r})e^{iq_1\mathbf{R}_n} + \rho_{1y}(\mathbf{r})e^{iq_2\mathbf{R}_n} + \rho_{1z}(\mathbf{r})e^{iq_3\mathbf{R}_n} \right) \\
	&+ b \left( \rho_{2x}(\mathbf{r})e^{iq_1\mathbf{R}_n} + \rho_{2y}(\mathbf{r})e^{iq_2\mathbf{R}_n} + \rho_{2z}(\mathbf{r})e^{iq_3\mathbf{R}_n} \right)	
	\end{split}
\end{equation*}
where $\rho_1$ and $\rho_2$ are active quadrupole functions while $\mathbf{R}_n$ define the $n$-th cubic unit cell of the crystal and $\mathbf{r}$ is the local radius vector, defining the electron density within the $n$-th cubic unit cell.

\begin{figure}
	\centering
\includegraphics[width=0.9\linewidth]{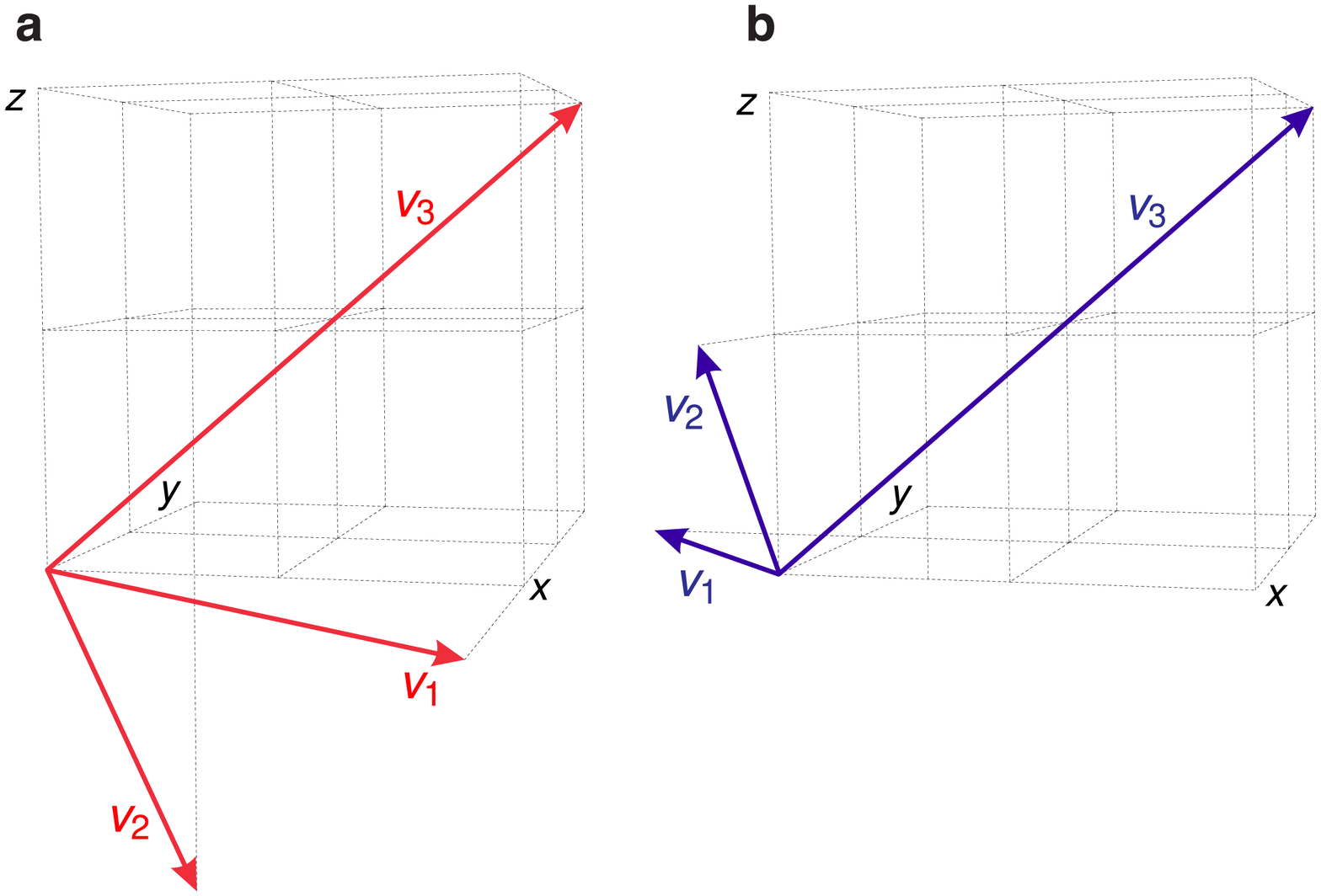}
\caption{Rhombohedral (hexagonal) unit cell of R3 ($\Gamma_\text{rh}$ $C_3^4$) for  the $X_1$ mode (a) and  $R_1R_3$ mode (b). $\mathbf{v}_1$ and $\mathbf{v}_2$ are perpendicular to $\mathbf{v}_3$. The angle between $\mathbf{v}_1$ and $\mathbf{v}_2$ is $2\pi/3$.}
\label{unitcell}
\end{figure}

The rhombohedral (trigonal) quadrupole phase has three basis vectors ($\mathbf{v}_1, \mathbf{v}_2, \mathbf{v}_3$), which in terms of the B20 cubic lattice vectors ($e_x, e_y, e_z$) are given by \cite{stokes88}
\begin{equation*}
		\mathbf{v}_1=2e_x-2e_y, \mathbf{v}_2=2e_y-2e_z, \mathbf{v}_3=2e_x+2e_y+2e_z.
\end{equation*}
The main (hexagonal) axis therefore is along the (1,1,1) main cube diagonal, see Fig.~\ref{unitcell}(a).

From the condensation scheme one can obtain invariant polynomials, which can be used to construct the Landau expansion of the free energy. Limiting ourselves to the polynomials of the fourth degree in order parameters, we obtain
\begin{align*}
\Phi(P,T,a,b) &= \Phi_0 + A (a^2+b^2) + B_1(a^2+b^2)^2 \\
			&+B_2(a^4+b^4) + B_3 a^2b^2 + B^4 ab (a^2-b^2).
\end{align*}
This expansion can be rewritten in the compact form in terms of the order parameter amplitudes $\eta$ and $\xi$ :
$$\Phi(P,T,a,b)=\Phi_0+A\eta^2+B_1'\eta^4 +B_2' \eta^4\sin2\xi \sin2(\xi+\xi_0)$$
Here $\eta=\pm \sqrt{a^2+b^2}$ and $\xi=\arccos(a/\eta)$, and $B_1'$, $B_2'$ are expressed through $B_1$, $B_2$, $B_3$, and $B_4$.
According to the Landau theory of the phase transitions above a certain critical temperature $T_Q$ which we call the quadrupole transition temperature, we have $A = \alpha (T - T_Q) > 0$, which leads to $\eta=0$. Below $T_Q$ the parameter $A$ becomes negative, which gives a nontrivial solution $\eta\neq 0$, $\xi_1\neq 0$, signaling the appearance of the new quadrupole phase. Notice that if we find one solution with $\xi_1$, we obtain the other solutions with the values $\xi_2=\xi_1+\pi/2$, $\xi_3=\xi_1+\pi$, $\xi_4=\xi_1+3\pi/2$. Different values of $\xi$ imply interchange between $\pm a$ and $\pm b$ with different signs (domains of R3). In terms of $\eta$ and $\xi_i$ (here $i$ = 1 or 2, 3, 4) we have
\begin{equation}
	a=\eta\cos(\xi_i-\xi_0), \; b=\eta\sin(\xi_i-\xi_0)	\label{eq1}
\end{equation}
From the expansion of the Landau free energy it follows that the phase transition would be of the second order if the Lifshitz condition \cite{lifshitz80} had been fulfilled. The Lifshitz condition requires that the coefficient $A$ in the expansion of the free energy as a function of $q$ should have a minimum for $q = q_1$ (and therefore for all other rays of the star, i.e. for $q = q_2$ and $q = q_3$). If the Lifshitz condition is violated, the Landau free energy depends on the quantity
$$ \eta\nabla\xi-\xi \nabla\eta \neq 0,$$
and the phase transition leads to the appearance of an incommensurate phase with a long wavelength modulation ($\lambda =2\pi/k\gg a$) of the order parameter amplitudes $\eta$ and $\xi$,
\begin{equation}
	\eta=\eta_0\cos(\mathbf{k}\cdot\mathbf{R}+\varphi_0), \; \xi=\xi_0\sin(\mathbf{k}\cdot\mathbf{R}+\varphi_0), \label{eq2}
\end{equation}
where $k\ll 1/a$. In this case, the macroscopic modulations of the order parameter amplitudes depend continuously on the thermodynamical parameters, in particular on temperature and pressure. From Eq.~(\ref{eq1}) we see that the variation of $\eta$ causes a gradual change of the quadrupole moments of Co, while the variation of $\xi$ implies a gradual rotation of the quadrupole component. Therefore, Eq.~(\ref{eq2}) describes a helical charge density wave, which is presumably directed along the (1,1,1) crystal axis.

The consideration of the $R_1R_3$ active mode at the $R$ point of the B20 Brillouin zone leads to the same qualitative conclusions. The $R$ point has one ray and the mode belongs to the four-fold irreducible representation $R_1R_3$ of the P2$_1$3 space group. The condensation scheme in the four dimensional space is given by the vector $(a,a,a,b)$. In real space new quadrupole electron density is given by
$$ \rho(\mathbf{R}_n,\mathbf{r})=[a\rho_1 (\mathbf{r}) + a\rho_2(\mathbf{r})+ a\rho_3(\mathbf{r}) + b \rho_4 (\mathbf{r})] e^{iq_R \mathbf{R}_n},$$
where $\rho_1 - \rho_4$ are four active basis functions. In this case three basis vectors are written as $\mathbf{v}_1=-e_x+e_y$, $\mathbf{v}_2=-e_y+e_z$, $\mathbf{v}_3=2e_x+2e_y+2e_z$.
The unit cell is smaller than for the $X_1$ mode and the trigonal axis is again along the (1,1,1) direction, see Fig.~\ref{unitcell}(b).

As before, the Landau free energy expansion in terms of the order parameter amplitude $a$ and $b$ is characteristic of a second order transition (since the Landau free energy for this case is quite complicated, we do not quote it here), but since the Lifshitz condition is violated, the transformation occurs to an incommensurate phase with a long wave-length modulation ($\lambda =2\pi/k \gg a$) of $a$ and $b$,
\begin{equation*}
	a=a_0\cos(\mathbf{k}\cdot\mathbf{R}+\varphi_0), \; b=b_0\sin(\mathbf{k}\cdot\mathbf{R}+\varphi_0),
\end{equation*}
This relation implies a rotational dependence of the quadrupole density about the vector $\mathbf{k}$ in the Brillouin zone.

\end{document}